\documentclass[preprint,showpacs,preprintnumbers,amsmath,amssymb,endfloa
ts*]{revtex4}
\usepackage{graphicx}

\begin{document}

\thispagestyle{empty}
\title{
Dependences of the van der Waals atom-wall interaction
on atomic and material properties
}
\author{
A.~O.~Caride,
G.~L.~Klimchitskaya,\footnote{On leave from
North-West Technical University,
St.Petersburg, Russia.}
V.~M.~Mostepanenko,\footnote{On leave from
Noncommercial Partnership ``Scientific Instruments'',
Moscow, Russia.}
and S.~I.~Zanette
}

\affiliation{
Centro Brasileiro de Pesquisas F\'{\i}sicas,
Rua Dr.~Xavier Sigaud, 150,\\
URCA, 22290--180, Rio de Janeiro, RJ, Brazil
}

\begin{abstract}
The 1\%-accurate calculations of the van der Waals
interaction between an atom and a cavity wall are performed
in the separation region from 3\,nm to 150\,nm. The cases
of metastable He${}^{\ast}$ and Na atoms near the metal,
semiconductor or dielectric walls are considered.
Different approximations to the description of wall material
and atomic dynamic polarizability are carefully compared.
The smooth transition to the Casimir-Polder interaction is
verified. It is shown that to obtain accurate results for the
atom-wall van der Waals interaction at shortest separations
with an error less than 1\% one should use the complete
optical tabulated data for the complex refraction index
of the wall material and the accurate dynamic polarizability of 
an atom. The obtained results may be useful for the theoretical
interpretation of recent experiments on quantum reflection 
and Bose-Einstein condensation of ultracold atoms on or near 
surfaces of different nature.
\end{abstract}
\maketitle

\section{Introduction}

The van der Waals interaction is the well known example of dispersion
forces and there is an extensive literature devoted to this
subject (see, e.g., monographs~\cite{1,2,3}).
These forces are of quantum origin and they become detectable with
a decrease of separation distances between atoms, molecules and
macroscopic bodies. Further miniaturization, which is the main
tendency of microelectronics, brings more and more attention to
the investigation of fine properties of the van der Waals
interaction.

The van der Waals force between an atom (molecule) and a cavity wall
has long been investigated. In Ref.~\cite{4} its interaction
potential was found in the form of
$V_3(a)=-C_3/a^3$ 
in nonrelativistic approximation ($a$ is the separation
between an atom and a wall). The coefficient $C_3$ was calculated and
measured for different atoms and wall materials, both metallic
\cite{5,6,7} and dielectric \cite{8,9}. 
The theoretical and experimental results were shown to be in
qualitative agreement. More precise measurements were performed in
Refs.~\cite{10,11}. Currently the van der Waals interaction attracts
considerable interest in connection with experiments on quantum
reflection of ultracold atoms on different surfaces \cite{12,13}.
With the increase of separation distances up to hundreds nanometers
and more to several micrometers, the relativistic and thermal effects
become significant changing the dependence of the van der Waals force 
on separation. At moderate separations up to 1\,$\mu$m for atoms
described by the static atomic polarizability near a wall made of
ideal metal at zero temperature the interaction potential was
found by Casimir and Polder \cite{14} in the form
$V_4(a)=-C_4/a^4$.

Both the van der Waals and Casimir-Polder interactions are of much
importance in connection with the experiments on Bose-Einstein
condensation of ultracold atoms confined in a magnetic trap near
a surface \cite{15,16,17}. They may influence the stability of
a condensate and the effective size of the trap \cite{17}. 
Conversely, the Bose-Einstein condensates can be used as  
sensors of the van der Waals and Casimir-Polder forces.
The presence of these forces leads to the shift of the
oscillation frequency of the trapped condensate \cite{18}.
Note that in application to ultracold atoms not their temperature
but the temperature of the wall is the characteristic parameter of 
the fluctuating electromagnetic field giving rise to the van der
Waals interaction \cite{18,19}.

It is common knowledge that the precision of frequency shift
measurements is very high. Interpretation of these measurements
requires accurate theoretical results for the van der Waals and
Casimir-Polder interaction beyond the expressions given by the 
simple asymptotic formulas (in fact coefficients $C_3$ and $C_4$ are not
constants but depend on both separation distance and temperature,
and there is smooth joining between the formulas at some
intermediate separations). In the case of the Casimir-Polder
forces such results were obtained in Ref.~\cite{19} for different
atoms near a metal wall with account of finite conductivity of
a metal, dynamic atomic polarizability and nonzero temperature.
In Ref.~\cite{18} the influence of the Casimir-Polder force
between Rb atoms and sapphire wall onto the oscillations of a 
condensate was investigated.

In the present paper we find accurate dependences of the
van der Waals atom-wall interaction on the dynamic polarizability
of an atom and conductivity properties of wall material.
As an example, two different atoms are considered (metastable
He${}^{\ast}$ and Na), and metallic (Au), semiconductor (Si)
and dielectric (vitreous SiO${}_2$) walls. All calculations are
performed within the separation distances 
$3\,\mbox{nm}\leq a\leq 150\,$nm (for Au at larger separations
the accurate theoretical results for the Casimir-Polder interaction were
obtained in Ref.~\cite{19}). The theoretical formalism for the
exact computation of the van der Waals interaction is given
by the Lifshitz formula \cite{20,21,22} adapted for the
configuration of an atom near a wall. 

At small separations,
characteristic for the van der Waals force, it is necessary to
use the complete optical tabulated data for the complex index
of refraction in order to find the behavior of the dielectric
permittivity along the imaginary frequency axis (at separations
$a\geq 150\,$nm, as was shown in Ref.~\cite{19}, the dielectric
function of the free electron plasma model can be used in the
case of an Au wall to find the Casimir-Polder interaction).
We compare the results obtained by the use of complete data
for the dynamic polarizability of an atom and the ones given
by the single-oscillator model. This gave the possibility to
obtain more accurate results than in Ref.~\cite{23}, where the
single-oscillator model was used for a hydrogen atom near a silver 
wall, and also to determine the accuracy of the single-oscillator
approximation for the dynamic polarizability in the calculations
of the van der Waals interaction. It is shown that to calculate
the atom-wall van der Waals interaction with an error less than
1\% at a separation of several nanometers both the complete optical
tabulated data of the wall material and the accurate atomic
dynamic polarizability should be used.

The paper is organized as follows. In Sec.~II we briefly present
the main formulas and notations for the van der Waals interaction
between an atom and a cavity wall. Sec.~III contains the accurate
theoretical results for van der Waals interaction of He${}^{\ast}$ 
and Na atoms with an Au wall. In Sec.~IV the analogical results are
presented for semoconductor (Si) and dielectric (vitreous
SiO${}_2$) walls. Sec.~V contains our conclusions and discussion.

\section{Lifshitz formula for van der Waals atom-wall interaction}

The Lifshitz formula for the free energy of atom-wall interaction
(wall is at a temperature $T$ at thermal equilibrium) can be
presented in the form \cite{19,22}
\begin{eqnarray}
&&{\cal{F}}(a,T)=-k_BT
\sum\limits_{l=0}^{\infty}
\left(1-\frac{1}{2}\delta_{l0}\right)
\alpha(i\xi_l)
\int_{0}^{\infty}k_{\bot}dk_{\bot}q_le^{-2aq_l}
\label{eq1} \\
&&\phantom{aaaaa}\times
\left\{2r_{\|}(\xi_l,k_{\bot})+
\frac{\xi_l^2}{q_l^2c^2}\left[r_{\bot}(\xi_l,k_{\bot})-
r_{\|}(\xi_l,k_{\bot})\right]\right\},
\nonumber
\end{eqnarray}
\noindent
where $\alpha(\omega)$ is the atomic dynamic polarizability,
$k_B$ is the Boltzmann constant, $\xi_l=2\pi k_BTl/\hbar$
are the Matsubara frequencies, $l=0,\,1,\,2,\,\ldots\,$, 
$\delta_{lk}$ is the Kronecker symbol, and
the reflection coefficients for two independent polarizations
of the electromagnetic field are 
\begin{eqnarray}
&&r_{\|}(\xi_l,k_{\bot})=
\frac{\varepsilon_lq_l-k_l}{\varepsilon_lq_l+k_l},
\nonumber \\
&&r_{\bot}(\xi_l,k_{\bot})=\frac{k_l-q_l}{k_l+q_l}.
\label{eq2}
\end{eqnarray}
\noindent
In Eqs.~(\ref{eq1}) and (\ref{eq2})
the notations 
\begin{equation}
q_l=\sqrt{k_{\bot}^2+\frac{\xi_l^2}{c^2}},\quad
k_l=\sqrt{k_{\bot}^2+\varepsilon_l\frac{\xi_l^2}{c^2}}
\label{eq3}
\end{equation}
\noindent
are also introduced,
where $\varepsilon_l=\varepsilon(i\xi_l)$ is the dielectric
permittivity computed at the imaginary Matsubara frequencies,
$k_{\bot}$ is the wave vector in the plane of the wall.

We will apply Eq.~(\ref{eq1}) in the separation region
$3\,\mbox{nm}\leq a\leq 150\,$nm which corresponds to the
van der Waals interaction (near the left-hand side of the interval)
and transition domain to the Casimir-Polder interaction. 
In fact, in this region
at room temperature $T=300\,$K the temperature effect is negligible.
For the sake of convenience in numerical computations we, however,
do not make the approximate change of the discrete summation for
integration over continuous frequencies and use the original exact
Eq.~(\ref{eq1}).

For further application in computations, we introduce
the dimensionless variables
\begin{equation}
y=2aq_l,\quad\zeta_l=\frac{2a\xi_l}{c}\equiv\frac{\xi_l}{\omega_c},
\label{eq4}
\end{equation}
\noindent
where $\omega_c\equiv\omega_c(a)=c/(2a)$ 
is the characteristic frequency of the 
van der Waals interaction.

Separating the zero-frequency term, Eq.~(\ref{eq1}) can be
represented in the form
\begin{eqnarray}
&&{\cal{F}}=-\frac{C_3(a,T)}{a^3},
\label{eq5} \\
&&
C_3(a,T)=\frac{k_BT}{8}\left\{
2\alpha(0)\frac{\varepsilon(i0)-1}{\varepsilon(i0)+1}+
\sum\limits_{l=1}^{\infty}
\alpha(i\zeta_l\omega_c)\right.\nonumber \\
&&\phantom{aa}\times\left.
\int_{\zeta_l}^{\infty}dye^{-y}
\left[\vphantom{\sum}
2y^2r_{\|}(\zeta_l,y)+
\zeta_l^2\left[r_{\bot}(\zeta_l,y)-
r_{\|}(\zeta_l,y)\right]\right]
\vphantom{\sum\limits_{l=1}^{\infty}}
\right\}.
\nonumber
\end{eqnarray}
\noindent
Note that for metal 
$[\varepsilon(i0)-1]/[\varepsilon(i0)+1]=1$, 
whereas for dielectrics and semiconductors this ratio is
equal to $(\varepsilon_0-1)/(\varepsilon_0+1)$, where 
$\varepsilon_0$ is the static dielectric permittivity.

In terms of the new variables the reflection coefficients (\ref{eq2})
are
\begin{eqnarray}
&&r_{\|}(\zeta_l,y)=
\frac{\varepsilon_ly-
\sqrt{y^2+\zeta_l^2\left(\varepsilon_l-1\right)}}{\varepsilon_ly+
\sqrt{y^2+\zeta_l^2\left(\varepsilon_l-1\right)}},
\nonumber \\
&&r_{\bot}(\zeta_l,y)=
\frac{\sqrt{y^2+\zeta_l^2\left(\varepsilon_l-1\right)}
-y}{\sqrt{y^2+\zeta_l^2\left(\varepsilon_l-1\right)}+y}.
\label{eq6}
\end{eqnarray}

In a nonrelativistic limit Eq.~(\ref{eq5}) leads to
\begin{equation}
C_3(T)=\frac{k_BT}{4}\left[
\alpha(0)\frac{\varepsilon(i0)-1}{\varepsilon(i0)+1}+
2\sum\limits_{l=1}^{\infty}
\alpha(i\xi_l)\frac{\varepsilon(i\xi_l)-1}{\varepsilon(i\xi_l)+1}
\right],
\label{eq7}
\end{equation}
\noindent
which gives the usual estimation for the value of the van der Waals
constant at the shorter separations. Remind that Eq.~(\ref{eq7})
practically does not depend on temperature. By using the
Abel-Plana formula \cite{24} it can be approximately represented by
\begin{equation}
C_3\approx\frac{\hbar}{4\pi}
\int_{0}^{\infty}
\alpha(i\xi)\frac{\varepsilon(i\xi)-1}{\varepsilon(i\xi)+1}
d\xi.
\label{eq8}
\end{equation}

In the next two sections Eqs.~(\ref{eq5})--(\ref{eq7}) will be
used for accurate calculations of the van der Waals force
between different atoms near the surfaces made of metallic,
semiconducting and dielectric materials.

\section{van der Waals interaction of H$\mbox{e}{}^{\ast}$ and 
N$\mbox{a}$ atoms with gold wall}

To calculate the van der Waals free energy of atom-wall interaction
one should substitute the values of the dielectric permittivity
of the wall material and dynamic polarizability of the atom at
imaginary Matsubara frequencies into Eqs.~(\ref{eq5}) and
(\ref{eq6}).

We consider the separation distances $a\leq 150\,$nm (at larger
separations the analytical representation for 
{${\cal{F}}$} was obtained in Ref.~\cite{19} using the plasma model
dielectric function and the single oscillator model for the
atomic dynamic polarizability; the agreement up to 1\% with
the results of numerical computations was achieved).
As a lower limit of separations under consideration we fix
$a=3\,$nm. At smaller separation distances there are additional
physical phenomena, connected with the atomic structure of a wall
material, which are not taken into account in Eq.~(\ref{eq5})
but can influence atom-wall interaction.
The most important of them are the repulsive exchange potentials
with a range of action up to a few angstr\"{o}ms, and the
spatially nonlocal interaction due to the surface-plasmon charge
fluctuations. The latter contributes essentially at separations
of the order of $v_F/\omega_p\sim 1\,${\AA}, where $v_F$ is
the Fermi velocity and $\omega_p$ is the plasma frequency
\cite{25}. As was proved in Ref.~\cite{25}, at much larger
separations (in fact, starting from $a\approx 3\,$nm) the usual
Lifshitz formula, given by Eqs.~(\ref{eq1}) and (\ref{eq5})
is already applicable.

Within the separation region under consideration the characteristic
frequency $\omega_c$ reaches and even exceeds (at the shorter
separations) the plasma frequency (for Au we use
$\omega_p=1.37\times10^{16}\,$rad/s \cite{26}). By this reason in our
case the plasma or Drude dielectric functions are not good
approximations for the dielectric permittivity in all relevant
frequency range and one should use the complete tabulated data for
the complex index of refraction for Au to calculate the imaginary
part of the dielectric permittivity Im$\varepsilon(\omega)$ along 
the real frequency axis. The dielectric permittivity along the
imaginary frequency axis is found by means of the dispersion
relation \cite{27}
\begin{equation}
\varepsilon(i\xi)=1+\frac{2}{\pi}
\int_0^{\infty}{\!}d\omega
\frac{\omega\,\mbox{Im}\varepsilon(\omega)}{\omega^2+\xi^2}.
\label{eq9}
\end{equation}
\noindent
The available tabulated data for Au extend from 0.125\,eV
to 10000\,eV ($1\,\mbox{eV}=1.519\times 10^{15}\,$rad/s).
At shorter separations, to obtain the values of the van der Waals
free energy correct up to four significant figures, one should
find the dielectric permittivity at first 1850 Matsubara
frequencies. Near the right border of the separation interval
($a=150\,$nm) it would suffice to use only 60--70 first
Matsubara frequencies. In fact, to obtain $\varepsilon$
by Eq.~(\ref{eq9}) with sufficient precision one should extend
the available tabulated data for the region $\omega<0.125\,$eV.
This is conventially done with the help of the
imaginary part of the Drude dielectric
function
\begin{equation}
\varepsilon(\omega)=1-\frac{\omega_p^2}{\omega(\omega+i\gamma)},
\label{eq10}
\end{equation}
\noindent
where $\gamma=0.035\,$eV is the relaxation frequency.
It should be reminded also that Eqs.~(\ref{eq1}), (\ref{eq2}),
(\ref{eq10}) are free from contradiction with the Nernst heat theorem
which arise when the Drude dielectric function is substituted into
the Lifshitz formula at nonzero temperature in the configuration of
two parallel plates made of real metal (see Refs.~\cite{28,29}
for more details).

The computational results for Au are presented in Fig.~1 where
$\log_{10}\varepsilon(i\xi)$ is plotted as a function of
$\log_{10}\xi$ starting from the first Matsubara frequency
(at $T=300\,$K one has $\xi_1\approx2.47\times 10^{14}\,$rad/s
and $\log_{10}\xi_1\approx 14.4$).

Other data to be substituted into Eq.~(\ref{eq5}) are the
values of the atomic dynamic polarizability at imaginary
Matsubara frequencies. The accurate data (having a relative error
of about $10^{-6}$) were taken from Ref.~\cite{30} for the atoms
of metastable He${}^{\ast}$ and from Ref.~\cite{31} for Na (see
also the graphical representation in Fig.~3 of Ref.~\cite{19}).
It is interesting to compare the values of $C_3(a,T)$ obtained
by the use of the highly accurate data for the atomic dynamic
polarizability and in the framework of the single oscillator
model
\begin{equation}
\alpha(i\zeta\omega_c)=
\frac{\alpha(0)}{1+\frac{\omega_c^2\zeta^2}{\omega_0^2}},
\label{eq11}
\end{equation}
\noindent
where for He${}^{\ast}$ it holds $\alpha(0)=315.63\,$a.u.,
$\omega_0=1.18\,$eV \cite{32} and for Na it holds
$\alpha(0)=162.68\,$a.u., $\omega_0=1.55\,$eV \cite{33}
(1\,a.u. of polarizability is equal to 
$1.48\times 10^{-31}\,\mbox{m}^3$).

The computational results for the van der Waals coefficient
$C_3$ in the case of Au wall versus separation are represented
in Fig.~2 for metastable He${}^{\ast}$ (a) and Na (b) by solid
lines. These lines are obtained by the use of the optical
tabulated data for Im$\varepsilon$ and accurate atomic
dynamic polarizability. In the same figure the long-dashed
lines show the results obtained with the same data for
Im$\varepsilon$ but with a single oscillator model (\ref{eq11})
for the atomic dynamic polarizability. The short-dashed lines
illustrate the dependence of $C_3$ on $a$ in the case of a wall
made of ideal metal but with the accurate atomic dynamic
polarizability.

As is seen from Fig.~2, the account of the realistic properties
of a wall metal is important at all separations under consideration.
At the shortest separation $a=3\,$nm the result for an ideal metal
differs from the accurate result given by the solid line by about
16\% for He${}^{\ast}$ and by 28\% for Na. These strong
deviations only slightly decrease with the increase of separation.

A few calculated results for the values of $C_3$ are presented
in Table~I at $T=300\,$K for different separations indicated in the
first column. In columns 2 and 3 the values of $C_3$ for
He${}^{\ast}$ atom are computed for ideal metal and by the use
of the optical tabulated data for Im$\varepsilon$, respectively,
and in both cases with an accurate atomic polarizability.
In column 4 the optical tabulated data for Im$\varepsilon$
were used in combination with the single oscillator model for
the atomic polarizability of He${}^{\ast}$. In column 5
the plasma model dielectric function was used in calculations
together with an accurate atomic polarizability of He${}^{\ast}$.
In columns 6--9 the calculational results for a Na atom
are presented in the same order.

As is seen from Fig.~2 and Table~I (columns 3 and 4), the use of
the accurate data for the atomic dynamic polarizability (if to
compare with the single oscillator model) is of most importance at
the shortest separations. Thus, at $a=3\,$nm the relative error of
$C_3$ given by the single oscillator model is 4.4\% for
He${}^{\ast}$ and 2.2\% for Na. At $a=15\,$nm the single oscillator
model becomes more precise. For He${}^{\ast}$ it leads to only
3.3\%, and for Na to 1.6\% errors.

It is interesting to compare the calculated results obtained by 
the use of the complete tabulated data for Im$\varepsilon$ of Au
and by the plasma model dielectric function [Eq.~(\ref{eq10})
with $\gamma=0$]. From columns 3 and 5 of Table~I for He${}^{\ast}$,
and 7 and 9 for Na one can conclude that the error, given by the
plasma model, decreases from 6.3\% for He${}^{\ast}$ and 10\% for
Na at $a=3\,$nm to 0.8\% for He${}^{\ast}$ and 1\% for
Na at $a=150\,$nm. This illustrates the smooth joining of our
present results for the van der Waals interaction obtained by the
use of the optical tabulated data for Au with the analytical results 
of Ref.~\cite{19} for the Casimir-Polder interaction found by the
application of the plasma model.

The nonrelativistic asymptotic values of $C_3$ can be calculated
by the immediate use of Eqs.~(\ref{eq7}) and (\ref{eq9}) combined
with the optical tabulated data for Im$\varepsilon$ and the
accurate atomic polarizability. This leads to the results
$C_3\approx 1.61\,$a.u. for He${}^{\ast}$ and 
$C_3\approx 1.37\,$a.u. for Na in rather good agreement with the
data of columns 3 and 7 of Table~I computed at the shortest
separation $a=3\,$nm. Note, however, that the asymptotic
values, achieved at separations $a<3\,$nm, may be already outside
of the application region of the used theoretical approach (see
discussion in the beginning of this section).

As was shown in Ref.~\cite{19}, the account of the atomic dynamic
polarizability strongly affects the value of the Casimir-Polder
interaction if to compare with the original result \cite{14}
obtained in the static approximation. We emphasize that in the
case of the van der Waals interaction the influence of dynamic
effects is even greater than in the Casimir-Polder case. Thus,
if we restrict ourselves by only static polarizability of
He${}^{\ast}$ atom, the values of $C_3$ are found to be 11.6
and 1.64 times greater than those given in column 3 of
Table~I at separations $a=3\,$nm and $a=150\,$nm, respectively.

\section{van der Waals interaction of H$\mbox{e}{}^{\ast}$ and 
N$\mbox{a}$ atoms
with semiconductor and dielectric walls}

In this section we apply the formalism of Sec.~II to find the
accurate separation dependences of the van der Waals interaction
between He${}^{\ast}$ and Na atoms and Si or vitreous SiO${}_2$
wall. The chosen separation interval 
$3\,\mbox{nm}\leq a\leq 150\,$nm is the same as in Sec.~III.
In the case of dielectric and semiconductor surfaces there
are additional interactions due to the charged dangling bonds
at separations 1--1.5\,nm (see, e.g., Ref.~\cite{34}).
This is a further factor restricting the application of the
conventional theory of van der Waals forces at very short
distances.

The tabulated data for the complex refraction index of Si
extend from 0.00496\,eV to 2000\,eV \cite{26}. This permits
not to use any extension of data to smaller frequencies when
using Eq.~(\ref{eq9}) in order to find the dielectric
permittivity at all contributing imaginary Matsubara frequencies.
The computational results for Si are presented in Fig.~3a
where $\varepsilon(i\xi)$ is plotted as a function of
$\log_{10}\xi$ ($\xi$ is measured in rad/s). The static
dielectric permittivity of Si is equal to 
$\varepsilon_0=11.66$.

Substituting the obtained results for $\varepsilon(i\xi)$ and
also the data for the atomic dynamic polarizability of
He${}^{\ast}$ and Na (the same as in Sec.~III) into 
Eqs.~(\ref{eq5}) and (\ref{eq6}), one finds the dependences
of the van der Waals parameter $C_3$ on separation.
The results are shown in Fig.~4a (for He${}^{\ast}$) and
Fig.~4b (for Na). The solid lines are obtained by the use
of the accurate atomic dynamic polarizabilities, and the
long-dashed lines by using the single oscillator model
given by Eq.~(\ref{eq11}). The short-dashed lines are
obtained with the accurate dynamic polarizability but 
on the assumption that the dielectric permittivity does
not depend on frequency and is equal to its static value.
At the shortest separation $a=3\,$nm the error in $C_3$
due to the use of the static dielectric permittivity is
approximately 13\% for He${}^{\ast}$ and 24\% for Na.

In Table~II a few calculated values of $C_3$ at $T=300\,$K
are presented at separations listed in column 1.
In columns 2 and 3 the values of $C_3$ for He${}^{\ast}$  are
computed by the use of a static dielectric permittivity and
optical tabulated data for Im$\varepsilon$, respectively,
and in both cases with the accurate atomic dynamic
polarizability. In column 4 the data for Im$\varepsilon$
were used in combination with the single oscillator model 
for He${}^{\ast}$ dynamic polarizability. In columns 5--7
the same results for a Na atom are presented.

Table~II and Fig.~4 permit to follow the influence of atomic
and semiconductor characteristics onto the van der Waals force. 
Thus, comparing columns 3 and 4 we notice that the use of the
single oscillator model leads to 4.4\% error at $a=3\,$nm and
3.1\% error at $a=15\,$nm for the atom of metastable He${}^{\ast}$.
For the atom of Na these errors are 1.8\% and 1\%, respectively. 
With the increase of separation distance up to 150\,nm
the errors given by the single oscillator model decrease down to
0.4\% for He${}^{\ast}$ and practically to zero for Na.
This confirms that at larger separations the single oscillator
model is quite sufficient for calculations of the van der Waals
interactions with  errors below 1\%.

Now let us consider the case of a dielectric wall 
(vitreous SiO${}_2$).
The tabulated data for the complex refraction index of SiO${}_2$
extend from 0.0025\,eV to 2000\,eV \cite{26}. This is also quite
sufficient to calculate the dielectric permittivity at all
contributing Matsubara frequencies by Eq.~(\ref{eq9}) with no
use of any extension of data. The dependence of 
$\varepsilon(i\xi)$ as a function of $\log_{10}\xi$ for
SiO${}_2$ is shown in Fig.~3b. The static dielectric permittivity
of SiO${}_2$ is equal to $\varepsilon_0=4.88$. 

The obtained results for $\varepsilon(i\xi)$ and
the data for the atomic dynamic polarizability of
He${}^{\ast}$ and Na are substituted into
Eqs.~(\ref{eq5}) and (\ref{eq6}).
The resulting  dependences of $C_3$ on separation
 are shown in Fig.~5a (for He${}^{\ast}$) and
Fig.~5b (for Na). As in Fig.~4, the solid lines are related
to the use
of the accurate dynamic polarizabilities, the
long-dashed lines to the single oscillator model,
and the short-dashed lines to the use of the static
dielectric permittivity and an
accurate dynamic polarizability.

Table~III, containing a few calculated results, is organized 
in the same way as Table~II related to the case of a semiconductor 
wall. It permits to find errors resulting from the use of the
static dielectric permittivity instead of the accurate dependence
of $\varepsilon(i\xi)$ on frequency, and a single oscillator
model instead of an accurate dynamic polarizability for 
the atom near the dielectric wall. Thus, at $a=3\,$nm the use 
of the static dielectric permittivity instead of the optical
tabulated data leads to 78\% error in the value of the 
van der Waals coefficient $C_3$ for He${}^{\ast}$ and to 95\%
error for Na. These errors decrease to 2.1\% and 6.9\%,
respectively, if one uses the dielectric permittivity
$\tilde{\varepsilon}\approx 2.13$ corresponding not to 
the zero frequency
but to the frequency region of visible light. With the use
of $\tilde{\varepsilon}$ the largest errors in the value
of $C_3$ are achieved, however, not at the shortest separation
but at the largest separation considered here 
$a=150\,$nm (15\% for He${}^{\ast}$
atom and 12.7\% for Na atom). At this separation the use of the
static dielectric permittivity $\varepsilon_0$
leads to 56.6\% error (for
He${}^{\ast}$) and 62\% error (for Na).

By the comparison of columns 3 and 4 in Table III we conclude
that at a separation $a=3\,$nm the use of the single oscillator
model results in 5\% error for He${}^{\ast}$ atom and in 3\%
error for Na atom. At $a=15\,$nm the corresponding errors are
3.6\% and 1.2\%, respectively. At a separation $a=150\,$nm
the errors due to the use of the single oscillator model are
0.6\% for He${}^{\ast}$ atom and practically zero for Na atom,
i.e., the single oscillator model is sufficient.

\section{Conclusions and discussion}

In the foregoing we have performed accurate calculations of the
parameter $C_3$ describing the van der Waals atom-wall
interaction for the atoms of metastable He${}^{\ast}$ and Na
near metallic, semiconductor and dielectric walls.
The separation region from 3\,nm to 150\,nm was considered
covering the proper nonrelativistic van der Waals interaction
and some part of the transition region to the relativistic
Casimir-Polder interaction. At $a=150\,$nm the smooth joining
of the obtained results with the calculations of Ref.~\cite{19}
for the Casimir-Polder case was followed.

It was shown that qualitatively the cases of an atom near
metallic, semiconductor and dielectric walls are very similar.
The use of approximations of the ideal metal or the static
dielectric permittivity leads to the errors in the value
of $C_3$ of about (13--28)\% at the shortest separation
depending on the wall material and the type of atoms. This error
slowly decreases with the increase of separation remaining
rather large in the case of metallic wall. The more adequate
(for metals) plasma model dielectric function results in
(6--10)\% errors at the shortest separation.

We have compared the results for $C_3$ obtained by the use of
the accurate atomic dynamic polarizability with those
obtained from the single
oscillator model. At the shortest separation the single oscillator
model leads to errors of about (1.8--4.4)\% in the values of
$C_3$. These errors quickly decrease with the increase of 
separation.

The magnitude of the error, given by one or another approximation
used, depends qualitatively on the type of the atom. By way of
example, for Na atom the use of a single oscillator model leads
to less errors than for He${}^{\ast}$ independently of wall
material.

The performed investigation permits to make a conclusion that the
accurate calculations of the van der Waals atom-wall interaction
at short separations with the error no larger than 1\% require the
use of both complete optical tabulated data of wall material and
accurate dynamic polarizability of an atom. This is distinct
from the case of the Casimir-Polder interaction with a metallic
wall which can be described with no more than 1\% error using
the plasma model dielectric function of a wall material and
the single oscillator model for the dynamic polarizability of
an atom.

The obtained results can be used for theoretical
interpretation of the experiments on quantum reflection and
Bose-Einstein condensation of ultracold atoms on (near)
surfaces of different nature, and also in investigation of 
other physical, chemical and biological processes, where
the precise information on the van der Waals and
Casimir-Polder forces is needed.

\section*{Acknowledgments}

The authors are grateful to J.~F.~Babb for stimulating
discussions and for giving accurate data on the atomic
dynamic polarizability of He${}^{\ast}$ and Na.
We gratefully acknowledge FAPERJ (Processes E--26/170.132
and 170.409/2004) for financial support.
G.L.K. and V.M.M. were partially supported by Finep.


\begin{figure*}
\vspace*{-7cm}
\includegraphics{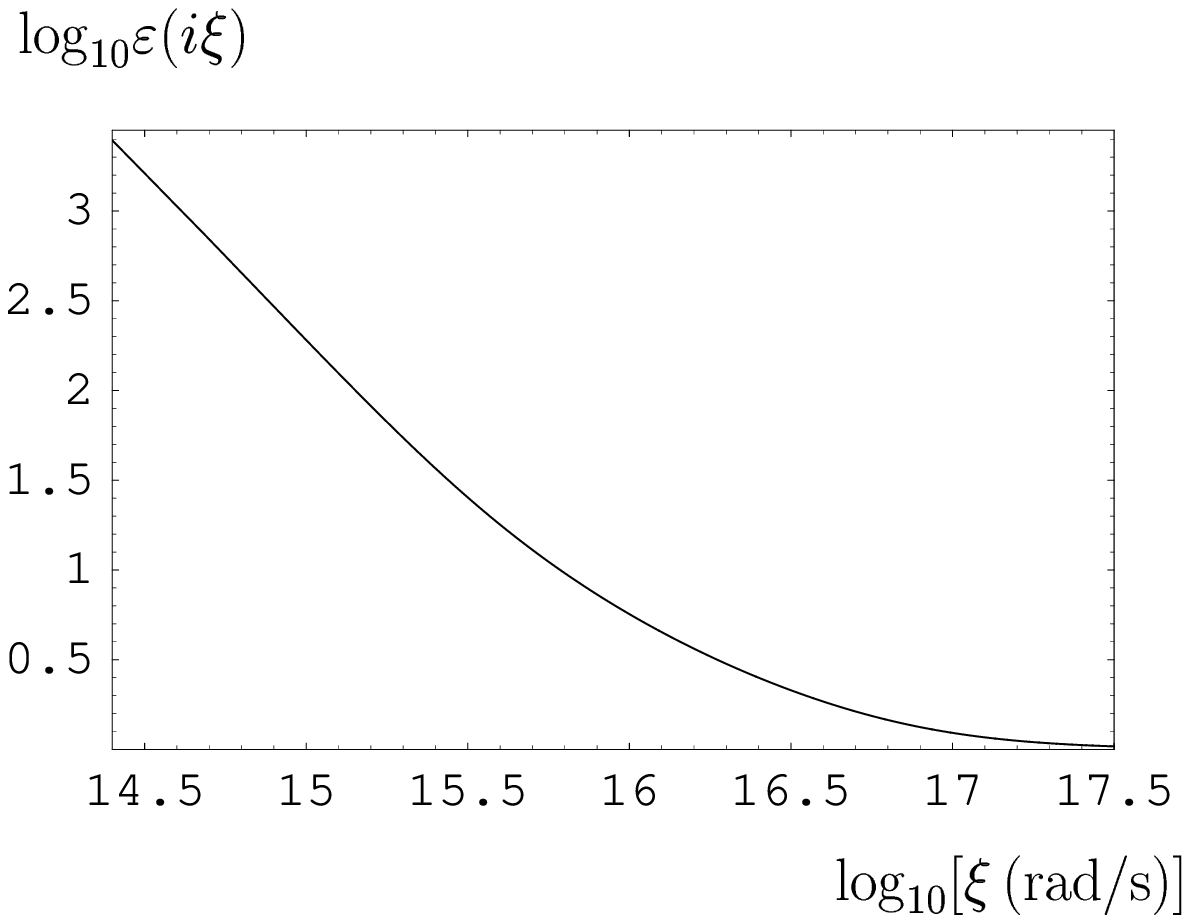}
\vspace*{-8.5cm}
\caption{
Logarithm of the dielectric permittivity of Au
along the imaginary axis as a function of the logarithm of
frequency.
}
\end{figure*}

\begin{figure*}
\vspace*{-4cm}
\includegraphics{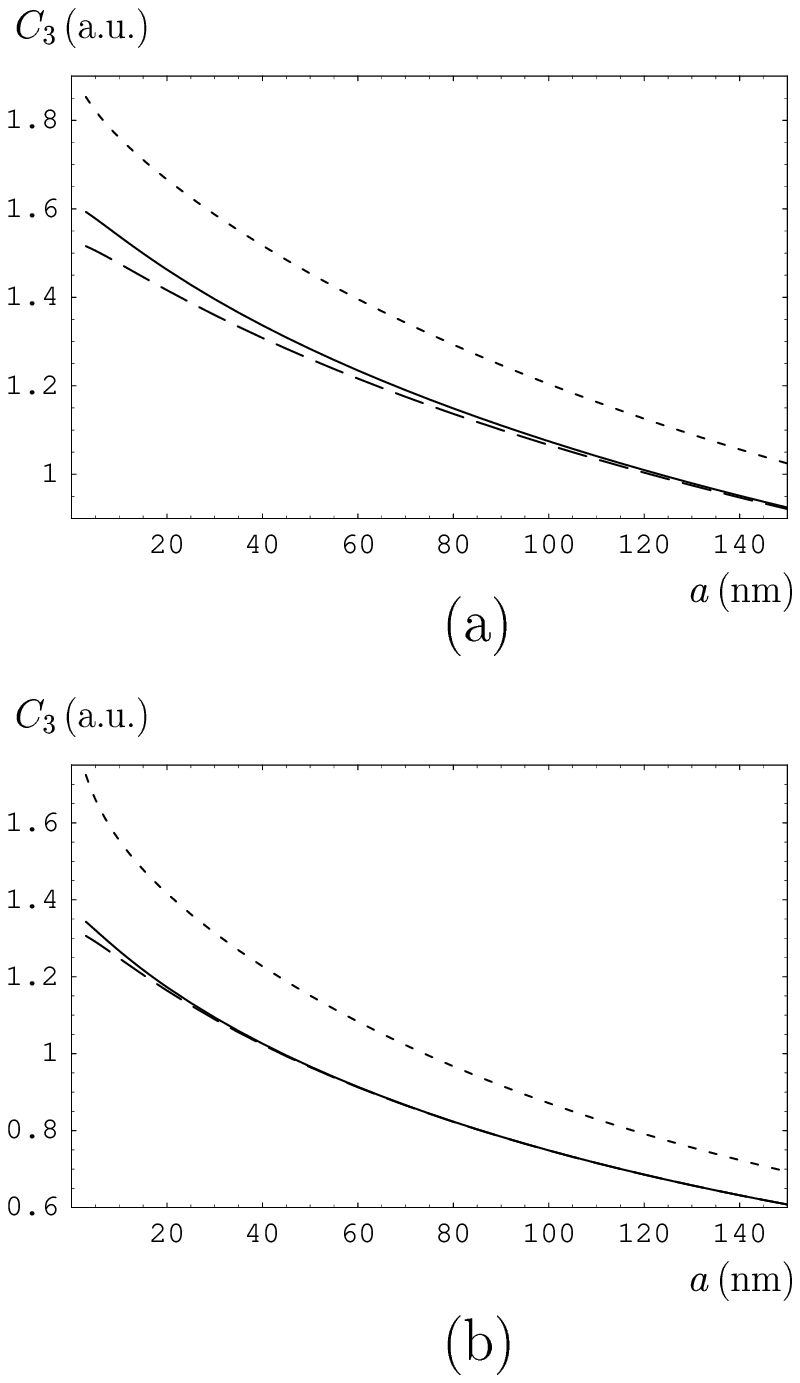}
\vspace*{-8.5cm}
\caption{
Dependence of the van der Waals coefficient $C_3$
on separation
for metastable He${}^{\ast}$ (a) and Na (b) atoms 
near Au wall calculated
by the use of the complete tabulated data of Au and
the accurate atomic dynamic polarizabilities (solid lines)
or by the single oscillator model (long-dashed lines). 
The short-dashed lines are calculated for the ideal metal 
with the accurate dynamic polarizability of the atoms.
}
\end{figure*}

\begin{figure*}
\vspace*{-4cm}
\includegraphics{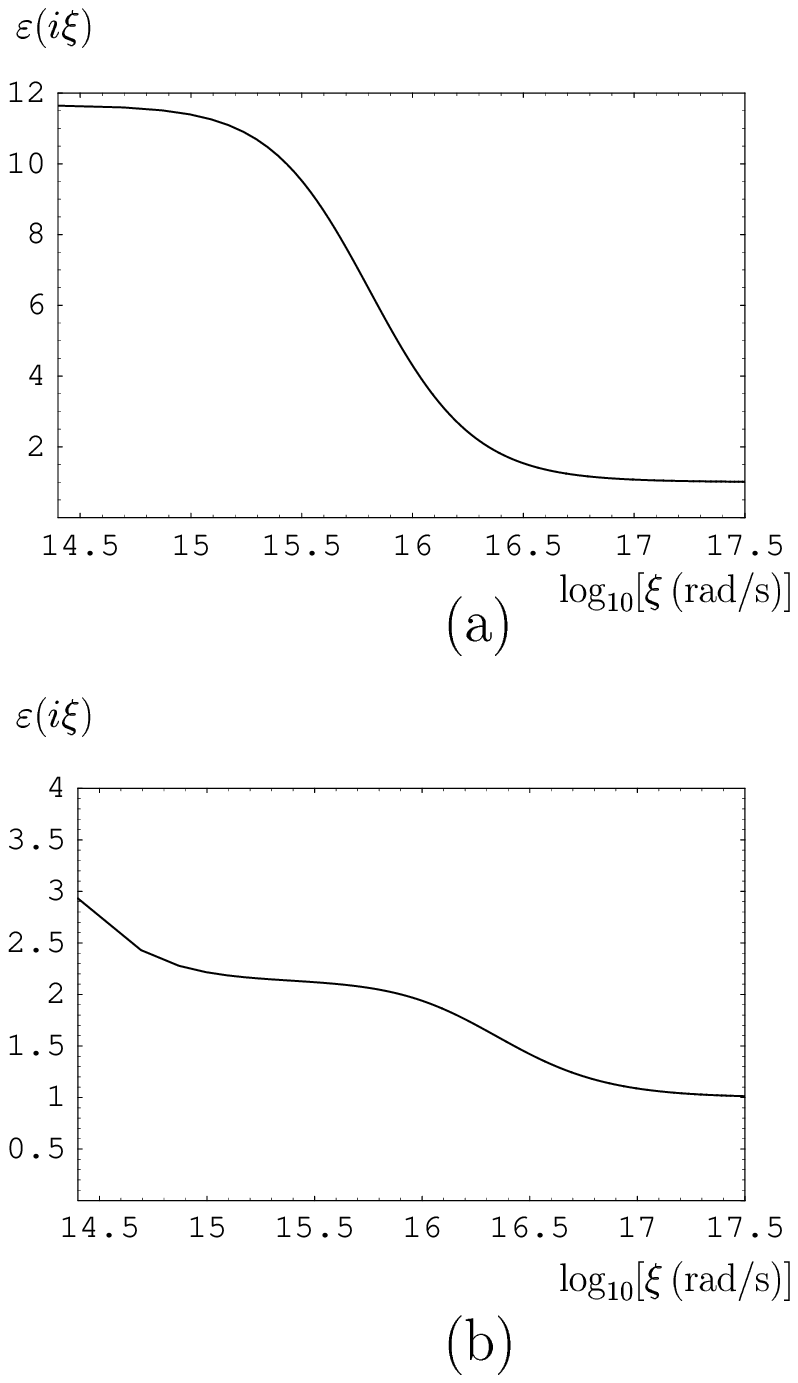}
\vspace*{-8.5cm}
\caption{
Dielectric permittivity of Si (a) and SiO${}_2$ (b)
along the imaginary axis as a function of the 
logarithm of frequency.
}
\end{figure*}

\begin{figure*}
\vspace*{-4cm}
\includegraphics{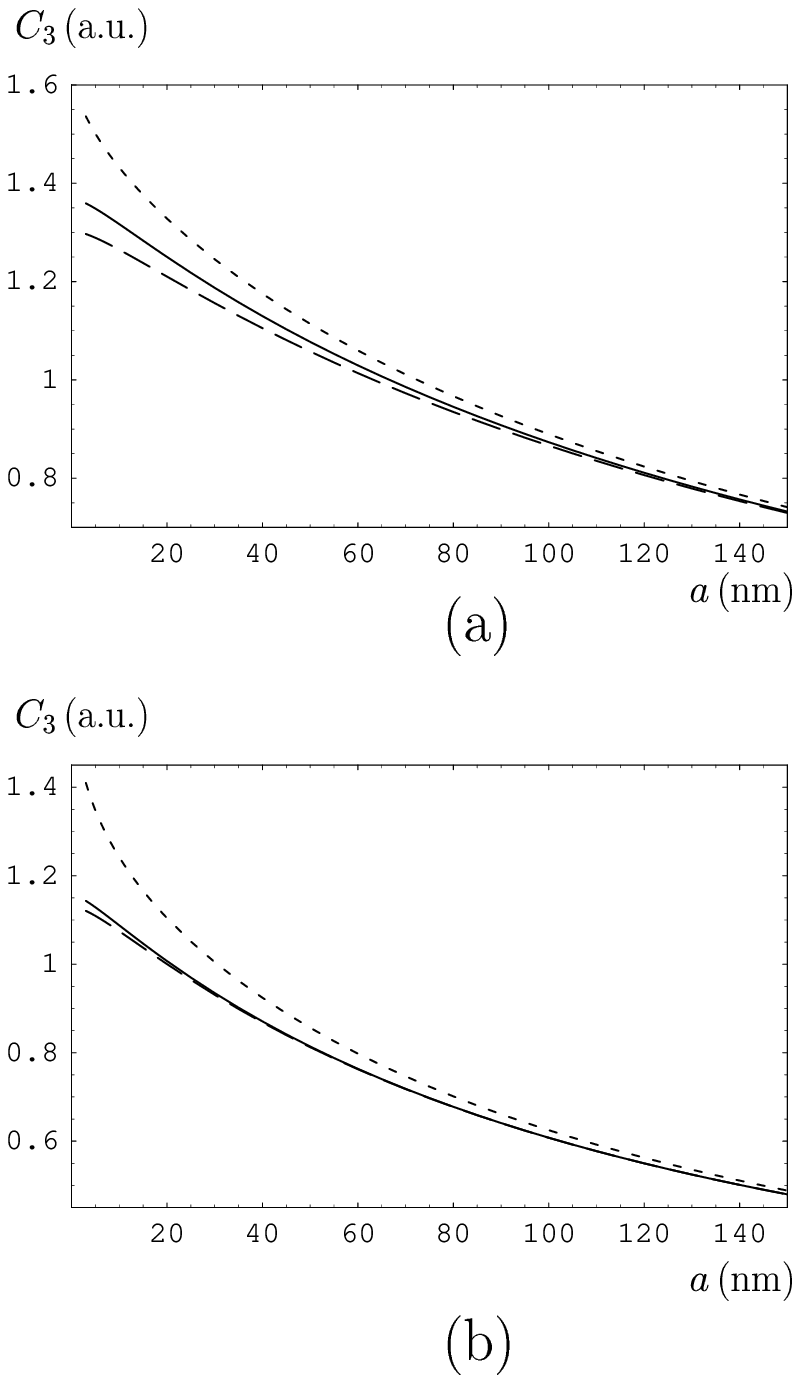}
\vspace*{-8.5cm}
\caption{
Dependence of the van der Waals coefficient $C_3$
on separation
for metastable He${}^{\ast}$ (a) and Na (b) atoms near
Si wall calculated
by the use of the complete tabulated data of Si with
the accurate atomic dynamic polarizabilities (solid lines)
and by the single oscillator model (long-dashed lines).
The short-dashed lines are calculated for semiconductor,
described by the static dielectric permittivity,
and by the accurate dynamic polarizability of the atoms.
}
\end{figure*}

\begin{figure*}
\vspace*{-4cm}
\includegraphics{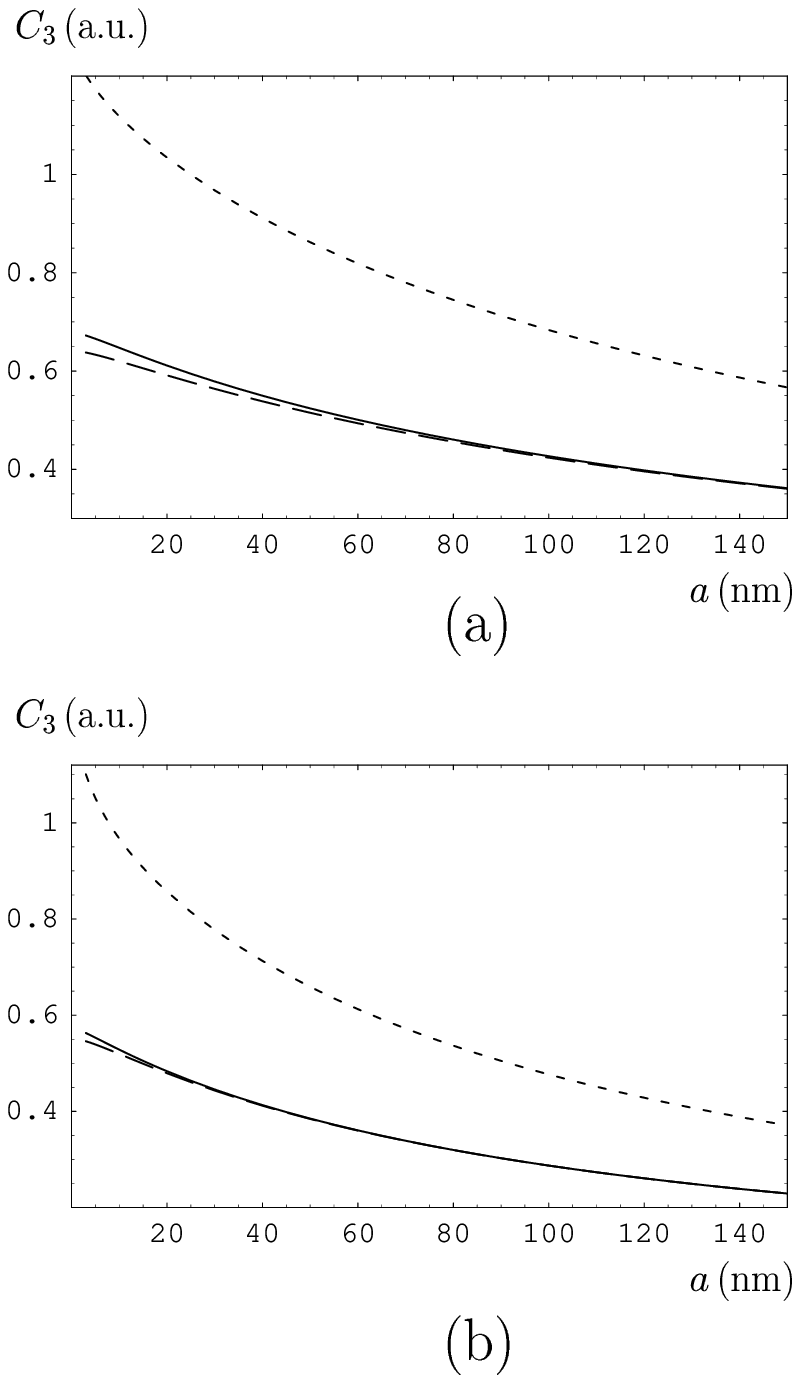}
\vspace*{-8.5cm}
\caption{
Dependence of the van der Waals coefficient $C_3$
on separation
for metastable He${}^{\ast}$ (a) and Na (b) atoms near
SiO${}_2$ wall calculated
by the use of the complete tabulated data of SiO${}_2$ with
the accurate atomic dynamic polarizabilities (solid lines)
and by the single oscillator model (long-dashed lines).
The short-dashed lines are calculated for dielectric,
described by the static dielectric permittivity,
and by the accurate dynamic polarizability of the atoms.
}
\end{figure*}
\begingroup
\squeezetable
\begin{table}
\caption{
Values of the coefficient $C_3$ of the van der Waals atom-wall
interaction at different separations computed for the ideal
metal (a) and for real metal (Au) described by the optical
tabulated data (b), and the accurate atomic dynamic 
polarizabilities; in column (c) real metal is described by the
optical tabulated data and the dynamic polarizability of an
atom is given by the single oscillator model; in column (d)
real metal is described by the plasma model and the dynamic
polarizability of an atom is accurate. 
}
\begin{ruledtabular}
\begin{tabular}{rcccccccc}
\multicolumn{1}{c}{$a$}
&\multicolumn{4}{c}{Metastable He${}^{\ast}$ near Au wall} &
\multicolumn{4}{c}{Na near Au wall }\\
\multicolumn{1}{c}{(nm)} & (a) & (b)&(c)&(d)&(a)&(b)&(c)&(d) \\ 
\hline
3 & 1.85 & 1.59 & 1.52 & 1.49 & 1.72 & 1.34 & 1.31 & 1.20 \\
5 & 1.82 & 1.58 & 1.51 & 1.48 & 1.66 & 1.32 & 1.29 & 1.19 \\
10 & 1.76 & 1.54 & 1.48 & 1.46 & 1.55 & 1.27 & 1.25 & 1.16 \\
15 & 1.71 & 1.50 & 1.45 & 1.43 & 1.48 & 1.22 & 1.20 & 1.13 \\
20 & 1.67 & 1.46 & 1.42 & 1.40 & 1.42 & 1.17 & 1.16 & 1.10 \\
25 & 1.62 & 1.43 & 1.39 & 1.38 & 1.36 & 1.13 & 1.12 & 1.07 \\
50 & 1.45 & 1.28 & 1.26 & 1.25 & 1.15 & 0.967 & 0.965 & 0.932 \\
75 & 1.32 & 1.17 & 1.16 & 1.15 & 0.994 & 0.844 & 0.844 & 0.822 \\
100 & 1.20 & 1.08 & 1.07 & 1.06 & 0.871 & 0.748 & 0.748 & 0.734 \\
125 & 1.11 & 0.994 & 0.989 & 0.985 & 0.773 & 0.671 & 0.671 & 0.662 \\
150 & 1.02 & 0.925 & 0.922 & 0.918 & 0.693 & 0.608 & 0.608 & 0.601 
\end{tabular}
\end{ruledtabular}
\end{table}
\endgroup
\begingroup
\squeezetable
\begin{table}
\caption{
Values of the coefficient $C_3$ of the van der Waals atom-wall
interaction at different separations computed for the semiconductor
(Si) described by the static dielectric permittivity
(a) and  by the optical
tabulated data (b), and the accurate atomic dynamic
polarizabilities; in column (c) semiconductor is described by the
optical tabulated data and the dynamic polarizability of an
atom is given by the single oscillator model.
}

\begin{ruledtabular}
\begin{tabular}{rcccccc}
\multicolumn{1}{c}{$a$}
&\multicolumn{3}{c}{Metastable He${}^{\ast}$ near Si wall} &
\multicolumn{3}{c}{Na near Si wall }\\
\multicolumn{1}{c}{(nm)} & (a) & (b)&(c)&(a)&(b)&(c) \\
\hline
  3 & 1.54 & 1.36 & 1.30 & 1.41 & 1.14 & 1.12 \\
  5 & 1.50 & 1.35 & 1.29 & 1.35 & 1.13 & 1.11 \\
 10 & 1.43 & 1.32 & 1.26 & 1.24 & 1.09 & 1.07 \\
 15 & 1.38 & 1.28 & 1.24 & 1.17 & 1.05 & 1.04 \\
 20 & 1.33 & 1.25 & 1.21 & 1.10 & 1.01 & 1.00 \\
 25 & 1.28 & 1.22 & 1.18 & 1.05 & 0.970 & 0.965 \\
 50 & 1.11 & 1.08 & 1.06 & 0.856 & 0.814 & 0.812 \\
 75 & 0.998 & 0.965& 0.954 & 0.723 & 0.698 & 0.697 \\
100 & 0.890 & 0.873 & 0.866 & 0.625 & 0.608 & 0.608 \\
125 & 0.809 & 0.797 & 0.792 & 0.549 & 0.537 & 0.537 \\
150 & 0.741 & 0.732 & 0.729 & 0.488 & 0.480 & 0.480
\end{tabular}
\end{ruledtabular}
\end{table}
\endgroup
\begingroup
\squeezetable
\begin{table}
\caption{
Values of the coefficient $C_3$ of the van der Waals atom-wall
interaction at different separations computed for the dielectric
(vitreous SiO${}_2$) described by the static dielectric permittivity
(a) and  by the optical
tabulated data (b), and the accurate atomic dynamic
polarizabilities; in column (c) semiconductor is described by the
optical tabulated data and the dynamic polarizability of an
atom is given by the single oscillator model.
}

\begin{ruledtabular}
\begin{tabular}{rcccccc}
\multicolumn{1}{c}{$a$}
&\multicolumn{3}{c}{Metastable He${}^{\ast}$ near SiO${}_2$ wall} &
\multicolumn{3}{c}{Na near SiO${}_2$ wall }\\
\multicolumn{1}{c}{(nm)} & (a) & (b)&(c)&(a)&(b)&(c) \\
\hline
  3 & 1.20 & 0.672 & 0.638 & 1.10 & 0.563 & 0.546 \\
  5 & 1.17 & 0.666 & 0.633 & 1.05 & 0.553 & 0.539 \\
 10 & 1.12 & 0.647 & 0.620 & 0.967 & 0.528 & 0.519 \\
 15 & 1.07 & 0.629 & 0.606 & 0.906 & 0.505 & 0.499 \\
 20 & 1.03 & 0.611 & 0.591 & 0.857 & 0.484 & 0.479 \\
 25 & 0.999 & 0.595 & 0.577 & 0.815 & 0.464 & 0.461 \\
 50 & 0.862 & 0.524 & 0.515 & 0.659 & 0.385 & 0.384 \\
 75 & 0.762 & 0.470 & 0.465 & 0.554 & 0.329 & 0.329 \\
100 & 0.684 & 0.427 & 0.424 & 0.477 & 0.287 & 0.287 \\
125 & 0.620 & 0.391 & 0.390 & 0.418 & 0.255 & 0.255 \\
150 & 0.567 & 0.362 & 0.360 & 0.371 & 0.229 & 0.229
\end{tabular}
\end{ruledtabular}
\end{table}
\endgroup
\end{document}